
%
%
%
%
\document
\magnification = 1200
\documentstyle{amsppt}

\baselineskip=16pt
\pageheight{19.2cm}
\pagewidth{12.9cm}
\hcorrection{0.5cm}
\topmatter
\title
Quantization of a class of piecewise affine transformations on the torus
\endtitle
\rightheadtext{Quantization of toral maps}
\author S. De Bi\`{e}vre${}^{\,(1)}$,
M. Degli Esposti${}^{\,(2)}$, R. Giachetti${}^{\,(3)}$,
\endauthor
\affil{
 ${}^{\,(1)}$ UFR de Math\'{e}matiques et
Laboratoire de Physique Th\'{e}orique et Math\'{e}matique
Universit\'{e} Paris VII - Denis Diderot 75251 Paris Cedex 05  France,\\
debievre\@mathp7.jussieu.fr\\ \\
${}^{\,(2)}$ Dipartimento di Matematica, Universit\`a di Bologna,\\
 Porta di Piazza San Donato 5, 40127 Bologna, Italy,\\
desposti\@dm.unibo.it\\ \\
${}^{\,(3)}$ Dipartimento di Fisica, Universit\`a di Firenze,\\
 Largo E. Fermi 2, 50125 Firenze, Italy,\\
giachetti\@fi.infn.it}
\endaffil
\thanks{The authors would like to thank Prof. Sandro Graffi for suggesting  the
problem
addressed in this paper and for many stimulating discussions. The first author
thanks the
department of mathematics of the University of Bologna, where part of this work
was performed
for its hospitality and the C.N.R. for partial financial support.}\endthanks
\date{(16 December 1994)}
\enddate
\abstract{ We present a unified framework for the quantization of a
family of discrete dynamical systems of varying degrees of ``chaoticity". The
systems to be quantized are piecewise affine maps on the two-torus, viewed as
phase space, and include the automorphisms, translations and skew translations.
We
then treat some discontinuous transformations such as the Baker map and
the sawtooth-like maps.
Our approach extends some ideas from geometric
quantization and it is both conceptually and calculationally
simple.}
\endabstract

\endtopmatter

\def\tende#1#2{\matrix \phantom{x}\\ \longrightarrow\\ {#1\rightarrow #2
               \atop\phantom{x}} \endmatrix}
\define\Lh{L^2 (\R^2,{dqdp \over 2\pi\hbar})}

\define\T{{\Bbb T}^{2}}

\define\R{\Bbb R}
\define\Q{\Bbb Q}
\define\C{\Bbb C}
\define\N{\Bbb N}
\define\Z{\Bbb Z}

\define\Qf{\widehat f}

\define\Qq{\widehat{q}}
\define\Qp{\widehat{p}}
\define\QB{\widehat{B}}
\define\Di{{\Cal S}'(\R^2)}
\define\Df{{\Cal S}_{\hbar}'(\theta)}
\define\Dfp{{\Cal S}_{\hbar}'(\theta')}

\define\eh{\frac{i}{2\hbar}}

\define\hw{{\Cal H}_w}
\define\hz{{\Cal H}_z}
\define\aw{{\Cal A}_w}
\define\dw{{\Cal D}_w}
\define\dz{{\Cal D}_z}
\define\dwt{{\Cal D}_w(\theta,N)}
\define\dzt{{\Cal D}_z(\theta,N)}
\define\hwt{{\Cal H}_w(\theta,N)}
\define\hzt{{\Cal H}_z(\theta,N)}
\define\hvt{{\Cal H}_v (\theta,N)}

\define\hwtp{{\Cal H}_w(\theta',N)}

\define\a1{\alpha_1}
\define\bt{\beta_{\theta}}
\define\qkw{q_k(w,\theta)}
\define\qlw{q_\ell(w,\theta)}

\define\g1{\gamma_1}
\define\ga2{\gamma_2}
\define\mt{\widetilde m}
\define\nt{\widetilde n}

\define\teti{\tilde{\theta}}

\magnification = 1200
\baselineskip=16 pt
\TagsOnRight
\nologo

\subheading{1. Introduction}\par
\bigskip

Interest in the quantization of discrete dynamical systems on compact phase
spaces comes from the desire to understand the
possible signature of classical chaotic dynamics in quantum mechanics.
Recall for example that it is expected and in some cases proved that the
asymptotic
properties ($\hbar\to 0$) of the eigenfunctions of quantized systems depend on
the degree of ``chaoticity'' of the corresponding classical ones (see, for
instance, \cite{Sar} and references therein). The torus forms an excellent
testing ground for these ideas. Indeed, the simplest ergodic
systems are the irrational translations on the torus, whereas the simplest
hyperbolic dynamical  systems are certain area-preserving maps \cite{AA,CFS}.
Among
these, the best known are the toral  automorphisms, the Baker transformation
and some discontinuous maps such as the sawtooth map considered in
\cite{Ch,LW,V,Li}. It has
been shown there that their singularities do not destroy the ergodicity and
mixing properties
one expects for hyperbolic maps.

One way in which the classical singularities will show up at the quantum level
is as follows.
For the linear automorphisms the classical and the quantum evolution are
identical, as in the
harmonic oscillator. The singularities will destroy this property, so that, to
control the
semiclassical behaviour of the eigenfunctions  a non trivial Egorov theorem
will be needed.
Similarly, the statistics of the  eigenvalues of the quantum propagator should
be more
generic than in the linear case, where they are determined by purely arithmetic
properties.
Clearly, before being able to address this kind of problems, one needs to
develop a
quantization for the systems considered.
Since none  of the above examples is obtained by evaluating a smooth
Hamiltonian flow
on the torus at discrete  times, the usual quantization schemes all fail and a
direct
attack is  needed.

 In this paper we will show how to extend the most elementary part of geometric
 quantization
\cite{Bl,GuSt, Ko, Sn, Wo} beyond its natural context in order to  construct a
unified and
simple  framework for the quantization of all of the above systems.  Some of
them
had not been quantized before, such as the translations and certain piecewise
affine hyperbolic
maps. It will turn out  that the  unitary matrices describing the quantum
evolution of each of those systems can be  computed straightforwardly and with
relatively
little effort in this way.

The toral automorphisms  and the Baker transformation were quantized
respectively in \cite{HB,DE, DGI} and in \cite{BV} and they have been studied
intensely ever since, both numerically and analytically
\cite{ Ke1, Ke2, Ke3, DGI, Eck, Sa}.  The methods of quantization used in these
papers look very different from each other.  Our approach reproduces the same
results
in those cases.

In order to get a more precise flavour of the ideas to be developed,
recall that in classical mechanics the dynamics of a system is obtained
by integrating a Hamiltonian vector field $X_H$ on a symplectic manifold
$(M,\omega)$. Here $H\in C^{\infty}(M)$ and $X_H$ is defined by
$X_H \rfloor\omega=dH.$ In quantum mechanics, the dynamics is given by a
unitary flow $U_t$ on a Hilbert space
${\Cal H}_{\hbar}$. A {\it quantization} is a set of rules allowing
to associate to $(M,\omega)$ a Hilbert space ${\Cal H}_{\hbar}$ and to
each function $f$ on $M$ in a suitable class ${\Cal C}$, a self-adjoint
operator  $\widehat f$ on ${\Cal H}_{\hbar}$.  One then says that
$U_t = \exp[\,(-i/\hbar)\widehat H t\,]$ is the quantization of the
classical flow of $X_H$. Typical requirements \cite{Be} are that the map
$f\mapsto \widehat f$ is linear, injective, unital, {\it i.e.} that it
satisfies
$\widehat 1 = Id_{{}\,{\Cal H}_{\hbar}}\,$, and that it is compatible with the
natural
involutions, $(\,\widehat f\,)^* = \widehat {\bar f\,}\,$.
Moreover, one requires the classical limit condition
$
(1/i\hbar)\,[\,\widehat f,\widehat g\,]\,\,\tende \hbar 0
\,\,\widehat {\{f,g\}}\,.
$

When the classical evolution is not a flow, but a discrete map, this scheme is
clearly not sufficient.
 We extend here some of the simplest ideas
of geometric quantization beyond their natural range of applicability
to obtain a unified framework for the quantization of
a reasonably large class of area preserving maps on the torus.  We will show
that, in spite of its reputation, the essence of geometric quantization is
intuitive, simple and well suited for such generalizations. For that purpose,
we first present in Section 2 a revisited version of the
geometric quantization on $T^*\R$, just to
demonstrate how it permits to {\it
reformulate} quantum mechanics for systems having $T^*\R$ as phase space
and to quantize linear flows. At
several points, we shall use physical or intuitive arguments to motivate parts
of the construction that are usually justified in terms of  very general
geometrical objects.
We then apply this approach to the quantization of toral automorphisms
in Section 3: the resulting quantum propagators are identical to the ones
obtained elsewhere by other methods \cite{HB,DE}.
In the final Section 4 we shall obtain the
quantization of translations, skew-translations as well as of
a class of {\it piecewise linear} hyperbolic maps such as the Baker
transformation and other maps studied, for instance, in \cite{Ch,LW,Li,V}.
Those
maps do not preserve the natural geometric structures associated with the
torus,
and therefore geometric quantization as such does not apply to them.
The proposed extension, however, will provide a definite answer.
\bigskip
\bigskip
\subheading{2. Geometric quantization on $T^*\R$}\par
\bigskip
As usual we call $(q,p)$ the coordinates of $T^*\R\cong\R^2$ and choose the
standard symplectic form $\omega = dq\wedge dp$ that gives the canonical
Poisson bracket
$\lbrace q,p \rbrace = 1\,$. Our goal is to realize the space of the quantum
states ${\Cal H}_{\hbar}$ as a subspace of $\Di$, equipped with a suitable
Hilbert space structure, and to establish a correspondence between classical
and quantum observables, so as to be able to describe the physical properties
of the quantum system. To this purpose we recall a first result, the validity
of which is easily checked by a direct computation: there exists a map
$f\in C^{\infty}(\R^2)\to\Qf\in L(\Di,\Di)\,,$ which is linear, unital
and satisfies the classical limit condition.
This map is explicitly given by
$$
\Qf=-i\hbar\nabla_{X_f} +f,\tag{2.1}
$$
where $X_f = (\partial_p f)\,\partial_q - (\partial_q f)\,\partial_p$
is the Hamiltonian vector field associated to $f$ and
$\nabla_X = X - (i/\hbar)\,X\rfloor \theta $
denotes the covariant derivative with respect to the connection form
$\theta = {1\over 2}(pdq - qdp)\,$.
Note that the use of $\nabla_X$ guarantees the local gauge invariance of
the construction (see \cite{Wo, Sn} for details).
It is moreover worth remarking that, if $\hat f$ in (2.1) is replaced
by $-i\hbar X_f$, then the unital property fails to hold, thereby
violating the uncertainty principle. In particular we have
$\widehat q = i\hbar\,\partial_p + q/2\,$ and
$\widehat p = -i\hbar \,\partial_q + p/2\,,$
so that, indeed, the canonical commutation relation
$\lbrack \widehat q, \widehat p \rbrack = i \hbar$ is satisfied.
The correspondence between $f$ and $\widehat f$ given in (2.1) is referred to
as {\it prequantization} \cite{Ko}.

We now need some conditions to choose the subspace ${\Cal H}_{\hbar}$ of $\Di$
and the Hilbert space structure it has to carry for it to correspond to the
quantum
Hilbert space of states. Note first that the equation
$i\hbar\partial_t\psi_t=\Qf\psi_t$
is easily solved on $\Di$. Writing $\psi_t=\exp{[\,-(i/\hbar)\widehat f
t\,]}\,\psi$,
one has
$$\align {}&\bigl(\exp[\,{\frac{i}{\hbar}\,\widehat f t}\,]\,\psi\bigr)(q,p)=\\
{}&{}\quad\quad\exp{[\,-\frac{i}{\hbar}\int_0^t ds\,
\bigl(\frac{1}{2}(p(s)\dot q(s)-q(s)\dot p(s))-f(q(s),p(s))\bigr)\,]}\,
\psi(q(t),p(t))\,.
\endalign
$$
where $(q(s),p(s))$ is the solution of the Hamilton equations
$\dot q=\partial_p f$, $\dot p=-\partial_q f$, with initial conditions
$(q,p)$.  Note that
the prequantized flow $\exp{[\,(i/\hbar)\widehat f t\,]}$ makes sense  also
when
$\psi\in\Di$.

The idea is then to try to pick ${\Cal H}_{\hbar}$ in such a way that
$\exp{[\,(i/\hbar)\widehat f t\,]}$ is a unitary one-parameter group for
a suitable large class ${\Cal C}$ of functions $f$.
This allows then for the interpretation of $\Qf$ as the quantized observable.

An a priori obvious choice would be $\Lh$.
It is nevertheless not suitable as the quantum Hilbert space.
Indeed it is easily seen that the spectra of $\Qq$ and $\Qp$ are not simple:
actually, the generalized eigenspaces are infinite dimensional, which is
in contradiction with standard quantum mechanics on
$L^2(\R)$. Otherwise stated, $\Qq$ (or $\Qp$) is not a complete set of
commuting observables on $\Lh$, or, equivalently, $\Qq$ and
$\Qp$ do not generate an irreducible algebra.
To put this more precisely, recall that the Heisenberg group is the group
$H=\R^3$ (as a set) equipped with the group law
$(a,b,\phi)(a', b', \phi') = (a+a', b+b', \phi +\phi'+{1\over 2}(ab'-a'b)).$
$H$ acts on $\R^2$ by $(a,b,\phi)(q,p)=(q+a, p+b)$. The
prequantized operators $\hat{q}$, $\hat{p}$  generate a unitary representation
of
$H$  on $\Lh$ given explicitly by
$$[U(a,b,\phi)\,\psi](q,p)
=\exp[\,{ - {i \over \hbar}\phi}\,]\,
\,\exp[\,{ -{i \over 2\hbar}(ap-bq)}\,]\,\,\,\psi(q-a,p-b).\tag{2.2}$$
This representation is not irreducible on $\Lh$.

There is a second problem with (2.1) which is worthwhile mentioning.
 It is easy  to see that, if $H(q,p) = p^2/2 + V(q)$, then
$\widehat H \not = \widehat p^{\,2}/2 + V(\widehat q)\,.$
It is then clear that the correspondence (2.1) is far from reproducing the
Schr\"odinger equation.

Some conditions have to be imposed on
the quantum Hilbert space ${\Cal H}_{\hbar}\subset \Di$ in order to
avoid the previous difficulties.
For the irreducibility of the algebra generated by $\hat q$ and $\hat p$ we
should require

$\phantom{ii}(i)$ ${} U(a,b,\phi)$ restricts to a unitary irreducible
representation of $H$ on ${\Cal H}_{\hbar}$.

\noindent To reproduce the Schr\"odinger equation we should impose:

$\phantom{i}(ii)$ ${}\quad\exists n_0\in\N^*$, and a dense subspace $D$ of
${\Cal H}_{\hbar}$  so that  $\widehat p$, ${\Qp}^{\,2}$, and ${\Qq}^{\,n}\,$
($1\leq n\leq n_0$) are essentially self-adjoint on $D$ and ${\Qp}^{\,2}=
\widehat{p^2\,}$,
${\Qq}^{\,n}=\widehat{q^n}$ on $D$.

\noindent Note that this is equivalent to requiring the correct form of
the Schr\"odinger equation for all polynomial potentials of order at most
$n_0$.  We are however already asking too much if we take $n_0\geq 2$, as we
now show.

\proclaim{Proposition 2.1} If $\psi\in\Di$ satisfies ${\Qp}^{\,2}\psi={\widehat
p}^2\psi$
and ${\Qq}^{\,2}\psi=\widehat{q^2}\psi$, then $\psi=0$.
\endproclaim
The proof of this proposition is a simple calculation that we omit.
In conclusion, the requirements $(i-ii)$  can not be satisfied on any non
trivial subspace
of $\Di$. Hence we can not even quantize in the proposed manner  Hamiltonians
with quadratic,
let alone general polynomial potentials.  The best we can still hope to do is
to impose
$(i)$ and a weakened version of $(ii)$, as we now explain.

Given $w\in\R^2$, with $w=(w_1,w_2)$, let $v\in\R^2$ such that $\omega(w,v)=1$,
we define the subspace
$${\Cal D}_w=\{\psi\in\Di \,\vert\, \nabla_w\psi=0\,\},\tag{2.3}$$
where $X_{h_w}$ is the Hamiltonian vector field associated to
$h_w(x)={}^T\!w\,x=w_1q+w_2p$ and $\nabla_w:\,=\nabla_{X_{h_w}}$. Here and
in the following $x\equiv (q,p)$. We then have
\proclaim{Lemma 2.1}
Let $w\in\R^2$ and $v\in\R^2$ such that $\omega(w,v)=1$. Then $\psi\in{\Cal
D}_w$ if and only if
there exists a tempered distribution  $f_v$ on the line such that
$$
\psi(x)=f_v(h_w(x))\, \exp[\,{-\frac{i}{2\hbar}\,h_w(x)h_v(x)\,]}.\tag{2.4}
$$
\endproclaim
\demo{Proof} We have
$\nabla_{w}=(w_2{\partial}_{q} -w_1{\partial}_{p})-(i/2\hbar)h_w(x)$.
Consider the map $(q,p)\mapsto (y_1,y_2)=(h_w(x),h_v(x))$ which is
linear and with determinant equal to unity. $\nabla_w\psi=0$  becomes
$\partial_{y_2}\,\eta(y_1,y_2)=-(i/2\hbar)\,y_1\,\eta(y_1,y_2)\,,$
with $\eta(y_1,y_2)=\psi(q,p)$. Its general solution is
$\eta(y_1,y_2)=f_v(y_1)\,\exp[\,{-(i/2\hbar)\,y_1y_2}\,]\,,$ thus
proving the lemma.
\qed\enddemo
\remark{Remark} If $v'\in\R^2$ satisfies $\omega(w,v')=1$, then $v'=v+rw$,
with $r\in\R$. It is easy to see that
$f_{v'}(y)=\exp[\,(i/2\hbar)ry^2\,]\,f_v(y)$.
We will therefore omit the indication of the dependence of $f$ on $v$.
We then have the following Lemma.
\endremark

\proclaim{Lemma 2.2} Let $\psi\in\Di$ and $w\in\R^2$. Then the following
are equivalent:
\roster
\item
$\widehat h{}_w^{\,2}\psi=\widehat{h_w^2}\psi$;
\item
$\widehat h{}_w^{\,n}\psi=\widehat{h_w^n}\psi$, for all $n\in\N$;
\item
$\psi\in\aw : =\{\eta\in\Di\,\vert\,(\nabla_{X_w})^2\eta=0\}$;
\item
Let $v\in\R^2$ such that $\omega(w,v)=1$.
Then there  exist $\varphi_0,\varphi_1\in{\Cal S}'(\R)$ such that
$$
\psi(x)=\bigl(h_v(x)\,\varphi_0(h_w(x)) +\varphi_1(h_w(x))\bigr)\,
\exp[\,{-\eh\,h_w(x)h_v(x)}\,]\,.
$$
\endroster \noindent
Moreover, if $u\in\R^2$, then $\widehat h_u\aw\subset \aw.$
\endproclaim
\demo{Proof}
A direct calculation shows
$
\widehat {h_w^n}=-i\hbar n\,h_w^{n-1}\nabla_{w} + h_w^n\,.
$  Using  $[\nabla_w,h_w(x)]=0$ to compare
$\widehat {h_w^n}$ to $(\widehat{h_w})^n$, and the previous lemma, the result
follows
easily.\qed
\enddemo

The lemma suggests to weaken $(ii)$ by imposing, $(\widehat{h_w})^n=\widehat
{h_w^n}$, for
some choice of $w$. This would imply $D\subset \aw$.
Now it is
not hard to see that the eigenvalues of $\hat q$ and $\hat p$ on $\aw$ are
doubly degenerate. In order to satisfy $(i)$ it would be natural to pick
$D$ in a subspace of $\aw$ on which this degeneracy is lifted. It is
 easy to describe all subspaces of $\aw$ that are, like $\aw$ itself,
invariant under all $\widehat h_u$, and on which the eigenvalues of all
$\widehat h_u$ are non-degenerate.  Although there seems to be no physical
 criteria permitting to select one such subspace, $\dw$ (see (2.3)) satisfies
the above
requirements and it is customary in geometric quantization to construct $\hw$
as a subspace
of $\dw$ because of its geometric appeal.
The condition $\nabla_w\psi=0$ is called a
{\it polarization condition} in this context. Note that we can identify
$\dw$ with
${\Cal S}'(\R)$ and that
$\widehat h_w$ then acts as a multiplication operator while $\widehat h_v$ as a
derivative operator.  A calculation as in the proof of Proposition $2.1$ shows
that
if $u\in\R^2$ is  not a multiple of $w$, then
$\widehat{h_u^2}\,\dw\cap\dw=\{0\}$, thus
excluding a priori the quantization of quadratic Hamiltonians, as already
pointed out.

Let us now briefly show how one can nevertheless correctly describe the
quantization of quadratic Hamiltonians within the framework of geometric
quantization (see [GuSt, Wo] for details).
Recall that for a quadratic polynomial
$f(q,p) = (\lambda/2)q^2 + \mu qp + (\nu/2)p^2\,$ the flow of
$X_f$ is linear and can be written as
${}^T\!(q(t),p(t))= A(t)\,{}^T\!(q,p)\,,$ with $A(t)\in SL(2,\R)$
(${}^T$ denotes the transpose).  The prequantized flow then reads
$$\bigl(\exp[\,{\frac{i}{\hbar}\,\widehat ft}\,]\,\psi\bigr)(q,p)
=\psi(A(t)\binom{q}{p})\,:\,= (U(A^{-1}(t))\psi)(q,p)$$
and the map $A\mapsto U(A)$ gives a unitary representation of $SL(2,\R)$
on $\Lh$. We now observe that $U(A)$ satisfies
$U(A)\dw={\Cal D_{{}\,{}^T\!\!A^{-1}w}}\,.$
We will explain below that it is possible to equip a suitable subspace $\hw$ of
$\dw$ with a
Hilbert space structure and then to identify the Hilbert spaces for different
values of $w$ by means of unitary maps $P_{zw}:\hw\rightarrow{\Cal H}_z$. This
is a particular case of a general construction  which allows to compare Hilbert
spaces corresponding to different real or complex polarizations (BKS kernels
\cite{Wo,GuSt, Sn}). The quantized linear transformation $V(A)$ is then
defined by $V(A)=D_\hbar P_{w,{}^T\!\!A^{-1}w}\circ U(A):\hw\to\hw$ (see
(2.8)).

We start by showing how to equip suitable subspaces $\hw$ of the $\dw$ with a
Hilbert space structure.   Note first that (2.4) implies that if
$\psi_1,\psi_2\in\dw$, then $\bar{\psi_1}\psi_2$ is a function of $h_w(x)$.
Moreover
$$
[\,\overline{U(a,b,\phi)\psi_1}\,U(a,b,\phi)\psi_2\,](q,p)=
\bar{f_1}f_2(h_w(x)-aw_1-bw_2).
$$
This suggests defining a Hilbert subspace $\hw$ of $\dw$ by
$$
\hw=\{\psi\in\dw\vert\,\int\vert\psi\vert^2(y)\,dy\,<\,\infty\},\tag{2.5}
$$
equipped with the obvious scalar product
$\langle\psi_1,\psi_2\rangle_{w}\,:\,=\int \bar{\psi_1}\psi_2(y)\,dy\,.$
The choice of the Lebesgue measure in (2.5) is dictated by the requirement
that $U(a, b,\phi)$ be unitary on $\hw$.

Let $w=(w_1,w_2)$ and $z=(z_1,z_2)$ be linearly independent and consider the
two
corresponding Hilbert spaces ${\Cal H}_{w}$ and ${\Cal H}_{z}$.
We denote by $v=(v_1,v_2)$ and $u=(u_1,u_2)$ two fixed vectors such that
$\omega(w,v)=\omega(z,u)=1$.
Consider $\psi\in{\Cal H_{w}}$ and $\varphi\in{\Cal H_{z}}$.
It is then easy to see that $\bar{\varphi}\psi$ belongs to $L^1(\R^2,dq\,dp)$.
The following proposition then follows from  a straightforward calculation that
we omit
\cite{GuSt}.

\proclaim{Proposition 2.2}
Let $w,z\in\R^2$ be linearly independent. Let $\Delta=\omega(w,z)$. Then there
exists a
unique continuous linear map $P_{zw}: {\Cal H_{w}}\to {\Cal H_{z}}$
such that, $\forall \psi\in {\Cal H}_w$ and $\forall \varphi\in {\Cal H}_z$
$$
\langle\varphi,P_{zw}\psi\rangle_{z}
=\int\bar{\varphi}\psi\,\frac{dq\,dp}{2\pi\hbar}\,. \tag{2.6}
$$
Moreover, if $D_\hbar\in\C$, with $\vert D_{\hbar}\vert=\sqrt{2\pi\hbar
\,\vert\Delta\vert\,}$\,, then $D_{\hbar}P_{zw}$ is unitary.
\endproclaim

The proof of the proposition provides an explicit expression for $P_{zw}$:
$$
(P_{zw}\,\psi)(x)=\frac{1}{2\pi\hbar\Delta}\,
\exp{[\,-\frac{i}{2\hbar}\,h_z(x)h_u(x)\,]}\,
\int f(y)\,\exp{[\,-\frac{i}{\hbar}S_{zw}(y,h_z(x))\,]}\, dy\,,
$$
where $S_{zw}$ is the quadratic form
$$
S_{zw}(y_1, y_2)=\frac{1}{2\Delta}\,\Bigl[\,(y_1,y_2)\left
(\matrix \omega(v,z) & 1\\ 1 & \omega(u,w)
\endmatrix\right)\binom{y_1}{y_2}\,\Bigr]\,.\tag{2.7}
$$
Note that $P_{zw}$ extends to $\dw$ (see \cite{Fo}).

The previous result allows to associate to any linear map $A\in SL(2,\R)$ and
to any given $z\in\R^2$ a well defined unitary operator, unique up to a phase,
in the following manner. Given $A\in SL(2,\R)$ and $z\in\R^2$, it follows
immediately that $\forall \psi\in\hz$ of the form (2.4), we have
$$
(U(A)\,\psi)(x)=f(h_{{}\,{}^T\!\!A^{-1}z}(x))\,
\exp{[\,-\eh\,h_{{}\,{}^T\!\!A^{-1}z}(x)\,h_{{}\,^T\!\!A^{-1}u}(x)\,]},
$$
where $U(A)$ is the previously defined prequantum action.
Hence $U(A)\hz={\Cal H}_{{}\,^T\!\!A^{-1}z}$ and we can define
$V(A):\hz\to \hz$ by
$$
V(A)=D_\hbar P_{z,{}^T\!\!A^{-1}z}\circ U(A).\tag{2.8}
$$
$V(A)$ is an unitary integral operator representing the quantum propagator
associated to the classical symplectic transformation $A$. Indeed, to see that
it agrees with Schr\"odinger quantum mechanics (up the the choice of a phase),
note that in the case where $z=(1,0)$ and $A=\left(\matrix a & b\\ c
& d \endmatrix\right)$, with $b\neq 0$, we recover the well known formula
($u=(0,1)$, $w=(d,-b)$, $v=(-c,a)$, $\Delta=b$) for the integral kernel
of $V(A)$, {\it i.e.}
$$
V(A)(y_1,y_2)=\frac{1}{\sqrt{2\pi\hbar b}}\,
\exp{[\frac{i}{2\hbar b}(ay_1^2-2y_1y_2+dy_2^2)\,]}\,.
$$
The correct phase for $V(A)$ is not obtained by the very simple approach
we have presented. This can be done with a considerable amount of
additional work \cite{Fo, GuSt}: this problem is however of no concern
in the present framework, since a global phase does not change the quantum
dynamics of a single transformation.
\bigskip
\bigskip

\subheading{3. Quantization of toral automorphisms}\par
\bigskip
We shall now apply our previous construction to quantization on the torus
$\T =\R^2/\Z^2$, with canonical  symplectic structure $\omega_{\T}$, such that
$d\pi^*\,\omega_{\T}=dq\wedge dp$, where $\pi:\R^2\to\T$ is the usual covering
map. In the first place, we need to identify the quantum Hilbert space.
The periodicity of the system in $q$ and in $p$ will be taken into account
along the same lines well known in solid state physics, namely by considering
distributions on $\R^2$ with quasiperiodic boundary conditions both in $q$ and
$p$. This approach is calculationally convenient and we shall show its
equivalence with the geometric quantization procedure.  It has the advantage of
being readily extendable to the geometrically non-natural situations of Section
4.

Let us introduce  $\,u_1=U(1,0,0)\,$ and $\,u_2=U(0,1,0)\,$ as in (2.2).
Given $\hbar\in \R^+$ and $\theta\in\T$, we denote by $\Df$ the space of all
the
tempered distributions $\psi$ on the plane satisfying the following conditions:
$$
u_1 \,\psi(q,p)=\exp[\,{2\pi i\,\theta_1}\,]\,\psi(q,p)\,,\quad\quad\quad
u_2 \,\psi(q,p)=\exp[\,{2\pi i\,\theta_2}\,]\,\psi(q,p)\,.\tag{3.1}
$$
Computing $(u_1u_2-u_2u_1)\psi$ using (2.2) and (3.1) one can easily see that
this space is non trivial if and only if
$
2\pi\hbar N=1
$
for some $N\in\N$. We shall refer to this as the  {\it prequantum condition}
and,
from now on, we shall assume it to be satisfied. In this case,
$\forall n=(n_1,n_2)\in\Z^2$ and $\psi\in\Df\,$,
$$
\psi(x+n)=\exp{[\,-2\pi i\,(\theta_1 n_1+\theta_2 n_2)\,]}\,
\exp{[\,\frac{i}{2\hbar}(qn_2-pn_1)\,]}
\exp{[\,-\frac{i}{2\hbar}n_1n_2\,]}\, \psi(x)\,,\tag{3.2}
$$
where, as in Section 2, $x=(q,p)$.
Given now $\psi(q,p)\in\Df$ and  $w\in\R^2$, one checks readily that
$\nabla_w \psi\in\Df$. We then define, in analogy with (2.3), the corresponding
space of linearly polarized sections $
\dwt=\Df\cap\dw=\{\psi\in\Df\vert\, \nabla_w\psi=0\}\,.
$
We will only consider polarizations of the torus for which $w_2/w_1\in\Q$.
This is equivalent to requiring that the flow lines of $X_w$ are circles.
In this case, up to rescaling $w$ by a constant multiple, we can
assume $w=(w_1,w_2)\in\Z^2$ with $g.c.d.\,(w_2,w_1)=1$.

\proclaim{Theorem 3.1}
Let $w=(w_1,w_2)\in\Z^2$ as above, where $g.c.d.(w_1,w_2)=1\,.$
Then  $\dwt$ is a complex vector space of dimension $N$.  Choosing $v\in\Z^2$
with $\omega(w,v)=1$, any $\psi\in\dwt$ can be written uniquely as
$$
\psi(x)=\sum_{k\in\Z}c_k\,\exp{[\,-i\pi N\, h_w(x)h_v(x)\,]}\,
\delta(h_w(x)-q_k(w,\theta))\,,\tag{3.3}
$$
where
$$
\qkw=k/N -(1/2)\,w_1w_2+(1/N)\,\omega(w,\theta)\,,\tag{3.4}
$$
and $\forall k\in\Z$, the $c_k\in\C$ satisfy
$$
c_{k+N}=\exp{[\,2\pi i\,\alpha_{\theta}(N,w)\,]}\,c_k\,,\tag{3.5}
$$
with
$$
\alpha_{\theta}(N,w)=(N/2)\,v_1v_2+\omega(v,\theta)\,.\tag{3.6}
$$
Conversely, any $\psi\in\Di$ of the form $(3.3)$ with the $c_k$'s satisfying
$(3.5-6)$ belongs to $\dwt$.
\endproclaim
\demo{Proof} Let $\psi\in\dwt$, then Lemma 2.1. implies that it is  of the form
$$
\psi(x)=f(h_w(x))\,\exp{[\,-i\pi N\,h_w(x)h_v(x)\,]},\tag{3.7}
$$
where $v\in\R^2$ is chosen such that $\omega(w,v)=1$.
It will be convenient to take $v\in\Z^2$. Note that, since
$g.c.d.\,(w_1,w_2)=1\,$, such $v$ always exists.
Using (3.2) and making the simple observation that
$$h_w(a)h_v(b)-h_v(a)h_w(b)=\omega(a,b)\,\quad\forall a,b\in\R^2\,,\tag{3.8}$$
one obtains, for any $n\in\Z^2$ and $t\in\R\,$, that $f$ must satisfy
$$
\align
f(t+h_w(n))=&\exp{[\,i\pi N\,(2th_v(n)+h_w(n)h_v(n))\,]}\\
{}&{}\quad\quad\exp{[\,-i\pi N\,n_1n_2\,]}\,
\exp{[\,-2\pi i\,h_n(\theta)\,]}\, f(t)\,.\tag{3.9}\\
\endalign
$$
Choosing $n=m$, where $m=(-w_2,w_1)$, and noting that
$h_w(m)=0\,$, $h_v(m)=\omega(w,v)=1\,$, $h_m(\theta)
=\omega(w,\theta)\,$,
one concludes that $f$ is of the form ($c_k\in\C$)
$$
f(t)=\sum_{k\in\Z} c_k\,\delta(t-q_k(w,\theta))\,,\tag{3.10}
$$
where the $q_k(w,\theta)$ are given in (3.4). Therefore (3.9) yields
$$
\sum_{k\in\Z} c_k\,\delta(t-q_k+h_w(n))= \sum_{k\in\Z}\,c_k\,\exp{[\,2\pi i\,
\bt (k,n,N)\,]}\,\delta(t-q_k)\,,
$$
where
$$
\bt(k,n,N)=Nq_k h_v(n)+\frac{N}{2}\,h_w(n)h_v(n)-\frac{N}{2}\,n_1n_2
-h_n(\theta)\,.\tag{3.11}
$$
Note that $\bt(k,n,N)$ depends on $k$ only through a term
$kh_v(n) =0\,\mod 1$.
Clearly we can drop this term and replace $\bt(k,n,N)$ by $\bt(n,N)$ defined by
$$
\bt(n,N)=-\frac{N}{2}w_1w_2h_v(n)+\omega(w,\theta)h_v(n)+
 \frac{N}{2}\bigl(h_w(n)h_v(n)-n_1n_2\bigr)-h_n(\theta).\tag{3.12}
$$
Observing that $q_k -h_w(n)=q_{k-Nh_w(n)}$ (see (3.4)), we find the following
condition on the $c_k$, $\forall n\in\Z^2$ and $\forall k\in\Z$:
$$
c_{k+Nh_w(n)}=\exp{[\,2\pi i\,\bt (n,N)\,]}\,c_k.\tag{3.13}
$$
We will show that the solution space of (3.13) is exactly $N$-dimensional.

First note that, for (3.13) to have any non-trivial solution at all, it is
necessary that
$$
\bt(n,N)=\bt(\nt,N)\,\mod 1\,,\tag{3.14}
$$
whenever $h_w(n)=h_w(\nt)$, {\it i.e.} whenever $\exists r\in\Z$
so that $\nt=n+rm$, ($m=(-w_2,w_1)$).
To prove (3.14), remark first that $\forall n,n'\in\Z^2$
$$
\bt(n+n',N)=\bt(n,N)+\bt(n',N)\,\mod 1\,.\tag{3.15}
$$
This follows immediately from (3.11) upon using  (3.8).
Moreover, one has
$$
\bt(m,N)=N\,[\,-\frac{1}{2}\,w_1w_2+\frac{1}{N}\,\omega(w,\theta)\,]
+\frac{N}{2}\,w_1w_2-\omega(w,\theta)=0\,\mod 1\,.
$$
This, together with (3.15) implies (3.14). We can then choose
$c_0,c_1,\cdots,c_{N-1}$ freely and
define $c_k$ for all other $k$ using (3.13).
To assure that the resulting solutions satisfy (3.13) $\forall k\in\Z$, and not
only for
$k=0,1,\cdots,N-1$, condition (3.15) is necessary and sufficient.

Finally, to compute $\alpha_{\theta}(N)$, note that
$\alpha_{\theta}(N)=\bt(n,N)$ for any $n\in\Z^2$ such that $h_w(n)=1$.
If we take $n=(v_2,-v_1)$, then $h_v(n)=0$, and (3.12) yields (3.6).\qed
\enddemo
\remark{Remark}
In particular, if $w=(1,0)$, it easy to see that the corresponding space
of polarized sections contains all distributions of the form
$f(q)=\sum_{k} c_k\,\delta(q-k/N-\theta_2/N)\,,$ where
$c_{k+N}=\exp{[\,-2\pi i\,\theta_1\,]}\,c_k$.
\endremark

Given now $w,v\in\Z^2$ and $\theta\in\T$ as before, the previous proposition
allows us to identify the space of sections $\dwt$ with $\C^N$, as follows:
$$
(c_0,\cdots,c_{N-1})\in\C^N \mapsto \psi(q,p)\in\dwt,\tag{3.16a}
$$
where
$$
\psi(q,p)=\sum_{k\in\Z} c_k\,\exp{[\,-i\pi N\, h_w(x)h_v(x)\,]}\,
 \delta (h_w(x)-\qkw).\tag{3.16b}
$$
Here, for $k\notin \{0,\cdots,N-1\}$, the $c_k$'s are defined by (3.5).

In analogy with the results of Section 2, we give $\dwt$ a Hilbert space
structure. Here also, the choice of the inner product will be dictated by
the requirement that the Heisenberg group acts unitarily. We shall denote by
${\Cal H}_w(\theta,N)$ the quantum Hilbert space thus obtained.

Setting $m=(-w_2,w_1)$ and $\mt=(v_2,-v_1)$ it is easy to see that $m$ and
$\mt$ form a basis of $\R^2$ and, in addition, that $\forall n\in\Z^2$, there
exist unique $\alpha,\beta\in\Z$ such that $n=\alpha m+\beta\mt$.
Moreover, by using (2.2), one  computes, for all
$\psi\in\dwt$ and for any $\alpha,\beta\in\R$ as
in (3.5),
$$
\align
\bigl(U(\alpha m)\psi\bigr)(x)&=\sum_{k\in\Z}[U(\alpha m)c]_k\,
\exp{[\,-i\pi N\, h_w(x)h_v(x)\,]}\,\,\delta(h_w(x)-q_k)\,,\\
\bigl(U(\beta\mt)\psi\bigr)(x)&=\sum_{k\in\Z}c_k\,
\exp{[\,-i\pi N\,h_w(x)h_v(x)\,]}\,\,\delta(h_w(x)-(q_k+\beta))\,,\\
\endalign
$$
where
$$
\bigl(U(\alpha m)c\bigr)_k=\exp{[\,2\pi iN\, q_k\alpha\,]}\,c_k\,.
$$
{}From these results and Theorem 3.1, we see that
$U(a,b) \dwt \subset \dwt$ if and only if $N(a,b) \in\ \Z^2$ and then
$$
\bigl(U(\frac{\ell}{N}\,m)\,c\bigr)_k=\exp[\,{2\pi i\, q_k\ell}\,]\,c_k\,,
\quad\quad\quad \bigl(U(\frac{\ell}{N}\,\mt)\,c\bigr)_k=
c_{k-\ell}\,.\tag{3.17}
$$
The natural Hilbert structure making $U(m)$ and $U(\mt)$ unitary is given by
$$
\langle\psi_2,\psi_1\rangle_w=
\sum_{k=0}^{N-1}{\bar{d}}_k c_k.\tag{3.18}
$$
where $\psi_1\cong (c_0,\cdots c_{N-1})$, $\psi_2\cong (d_0,\cdots d_{N-1})$.

As in section 2, we can construct a natural identification (or pairing) between
$\hwt$ and $\hzt$ when $w$ and $z$ are linearly independent.
  We first introduce the equivalent of the right hand side of (2.6).
  If $\psi_1\in\hwt$ and $\psi_2\in\hzt$, then $\bar{\psi_2}\psi_1$
  can be interpreted as a distribution on the plane. Indeed, although the
product of
  distributions is not defined in general, it makes sense in this case because
of the
  particular form of $\psi_1$ and $\psi_2$: $\delta(h_w(x)-q_l(w, \theta))$ and
  $\delta(h_z(x)-q_k(z, \theta))$ are supported on transversal lines, so that
we have
  no trouble defining their product.  Clearly, $\bar{\psi_2}\psi_1$  is
$\Z^2$-periodic
  and, as such, passes to a distribution on $\T$. Hence $\int_{\T}
  \bar{\psi_2}\psi_1\,\frac{dq\,dp}{2\pi\hbar}$ makes sense as the value of the
distribution
  $\bar{\psi_2}\psi_1$ on the function $(2\pi\hbar)^{-1}$ on $\T$.
  We then have, in analogy with Proposition 2.2:
\proclaim{Proposition 3.1}
  Given $w,z\in\Z^2$ as above with $\Delta=\omega(w,z)>0$ and  $\theta \in\T$,
there
  exist a unique vector space homomorphism
  $P_{zw}(\theta, N): \hwt\to \hzt$,
  such that $\forall \psi\in\hwt, \forall\varphi\in\hzt$
  $$
  \langle\varphi,P_{zw}(\theta, N)\psi\rangle_{\hzt}=\int_{\T}\bar
{\varphi}\psi\,
  \frac{dq\,dp}{2\pi\hbar}.\tag{3.19}
  $$
  Moreover, using the identifications defined in $(3.16)$, the matrix
representation of $P_{zw}(\theta, N)$ is
  $$P_{zw}(\theta, N)_{kr}=\frac{N}{\Delta}\,\sum_{p=0}^{\Delta-1}
\exp{[2\pi i\,\alpha_\theta (N,w)p]}\,
  \exp{[-2\pi iN\,S_{zw}
  (q_r(w,\theta)+p,q_k(z,\theta))]}.\tag{3.20}$$
  \endproclaim
\demo{Proof} That $P_{zw}$ is defined as a vector space homomorphism
by (3.19) is clear. To prove the rest of the proposition, we compute the right
hand side
of (3.19). Recall that this can be done by "integrating"
$\bar{\varphi}\psi$ over {\it any} fundamental domain of the torus. We start by
describing a suitable choice. Let $J=\left(\matrix 0 & 1\\
-1&0\endmatrix\right)$ and
define
$g_1=(1/\Delta)\, Jz,\quad g_2=-(1/\Delta)\, Jw$.
Then $g_1,g_2$ is a basis of $\R^2$ dual to $w,z$ since
$
h_w(g_1)=h_z(g_2)=1\,,\,\, h_w(g_2)=h_z(g_1)=0.
$
The unit cell of the dual lattice has volume $\Delta^{-1}$. Taking
$L=(L_1,L_2)\in\R^2$,
define
$$
T(L)=\{x\in\R^2\,\vert x=\alpha g_1+\beta g_2,L_1\leq\alpha<L_1+\Delta, L_2\leq
\beta<L_2+1\},\tag{3.21}
$$
which is the union of $\Delta$ dual unit cells. It is easy to see that
$T(L)$ is a fundamental domain for the torus. Indeed, suppose that $x=
\alpha g_1+\beta g_2$ and $x'=\alpha'g_1+\beta' g_2$ belong to
$T(L)$ and that $\exists n\in\Z^2$ so that $x'=x+n$. Then, (3.21) implies that
${}^T\!(\alpha'-\alpha,\beta'-\beta)= A(t)\,{}^T\!(h_w(n),h_z(n))\,.$
But $-1<\beta'-\beta<1$ and $h_z(n)\in\Z$, so $h_z(n)=0$, which implies that
$\exists
\gamma\in\R$ so that $n=\gamma (z_2,-z_1)$. Since $g.c.d.\,(z_1,z_2)=1$, it
follows that
$\gamma\in\Z$. Finally, this implies that $\alpha'-\alpha=h_w(n)=\gamma\Delta$
and, since
$-\Delta<\alpha'-\alpha<\Delta,\,\gamma=0$, so $ n=0$. We will use $T(L)$ with
a suitable
choice of $L$ to compute
$\int_{\T}\bar{\varphi}\psi\,\frac{dq\,dp}{2\pi\hbar}$.

For that purpose, recall that $\psi\in\hwt$ is supported on the lines
$h_w(x)=\qlw$, $\ell\in\Z$ and
$\varphi\in\hzt$ on the lines $h_z(x)=q_k(z,\theta)$,
$k\in\Z$, which intersect in the points $\{x_{k\ell}\,\vert k,\ell\in\Z\}$
defined
uniquely by
$
h_w(x_{k\ell})=\qlw, h_z(x_{k\ell})= q_k(z,\theta).
$
It is then clear that $ x_{k\ell}=\qlw g_1+q_k(z,\theta) g_2$, so that,
$\forall r,s\in\Z\,,$ $x_{k+rN,\ell}=x_{k\ell}+rg_2$ and
$x_{k,\ell+sN}=x_{k\ell}+sg_1$.
As a result, for a suitable choice of $L$
the points $x_{k\ell}$ belonging to $T(L)$ are
$
\{x_{k\ell}\,\vert\, 0\leq k< N-1,\,0\leq \ell< \Delta N-1\}.
$
Taking $\psi\cong (c_0,\cdots,c_{N-1})\in\C^N\cong \hwt$ and $\varphi\cong
(d_0,\cdots,d_{N-1})\in\C^N\cong \hzt$,(see (3.16))  we then readily obtain
that
$$
\align
&\int_{\T}\bar{\varphi}\psi\,
  \frac{dq\,dp}{2\pi\hbar}= \frac{N}{\Delta}\,
\sum_{k=0}^{N-1}\,\sum_{\ell=0}^{\Delta
   N-1}\overline{d}_kc_{\ell}
   \exp{[-2i\pi N S_{zw}(q_{\ell}(w,\theta),q_k(z,\theta))]}\\
&=\frac{N}{\Delta}\,\sum_{k=0}^{N-1}
\overline{d}_k
\sum_{r=0}^{N-1} c_r\sum_{p=0}^{\Delta -1}\exp{[2i\pi \alpha_{\theta}(N,w)p]}\,
\exp{[-2i\pi N S_{zw}(q_r(w,\theta)+p,q_k(z,\theta))]}\\
&=\sum_{k=0}^{N-1} \,\sum_{r=0}^{N-1}
\overline{d}_k P_{zw}(\theta, N)_{kr} c_r,
\endalign
$$
where we wrote $\ell=pN+r$ and used
$c_{pN+r}=c_r\,\exp{[2i\pi\alpha_{\theta}(N,w)p]}$, (see (3.5)). In conclusion,
the matrix representation of $P_{zw}(\theta)$
is given in (3.20).\qed\enddemo
The above definition of the pairing $P_{zw}(\theta, N)$ is a special case of a
very
general definition in the context of geometric quantization \cite{Sn}.  It
should be
remarked however
that the general theory does not guarantee  that the pairing is unitary: this
has to be
checked in each case separately.  We now turn to this task.  Note that the
explicit expression of the
matrix of $P_{zw}(\theta, N)$ is sufficiently complicated to make a direct
computation of
$P_{zw}^*(\theta, N)\,P_{zw}(\theta, N)$ difficult (except in the case when
$\Delta=1$, in
which case it is trivial).  We therefore develop a different
argument which uses the universal cover $\R^2$ of $\T$ and the known
unitarity of the pairing there (Proposition 2.2). This yields a proof for all
$P_{zw}(\theta, N)$ at once. It would be nice to have a direct geometric
proof for each fixed $\theta$.

\proclaim{Proposition 3.2}
\roster
\item
For any $w\in\Z^2$ with g.c.d.$(w_2,w_1)=1$,
$
\hw\cong N^2\,\int d^2\theta\,\hwt\,.
$
\item
$
U(a,b) P_{zw}=P_{zw}U(a,b)
$
for any $w,z\in\R^2$ and $\forall (a,b)\in\R^2\,.$
\item $P_{zw}= N\,\int\,d^2\theta\,P_{zw}(\theta, N)$.
Given $D_\hbar\in\C$, $\mid D_\hbar \mid =(\Delta/ N^3)^{1/2},$ \newline
$D_\hbar P_{zw}(\theta, N)$ is unitary.
\endroster
\endproclaim

\noindent \remark{Remark}
Note that if $w=(1,0)$, $z=(0,1)$ and $\theta=(0,0)$ then
$N^{-3/2}\,P_{zw}(\theta,N)={\Cal F}_N$. Here
${\Cal F}_N$ denotes the usual finite Fourier transform namely,
$$
f_\ell=\frac{1}{\sqrt N}\sum_{k=0}^{N-1}
\exp{[\,-\frac{2\pi i}{N}\,\ell k\,]}\,c_k\,.
$$
\endremark

 Before proving this proposition, note that the quantum map associated to any
$A\in SL(2,\Z)$ is now constructed exactly as in Section 2.
Let $A=\left(\matrix a & b\\ c&d\endmatrix\right)$ and take $\psi\in\Df$.
It is easy to see that $U(A) \psi= \psi\circ A^{-1}$ defines a map from  $\Df$
to $\Dfp$, where
$
\theta'=\beta(\theta)={}^T\!\!A\,\theta +\frac{1}{2}N\,{}^T\!(ac,db)\,\mod 1
$.
Moreover, for any $z\in\Z^2$ we have a natural map
$
 U(A): \hzt\to {\Cal H}_{{}\,{}^T\!\!A^{-1}z}(\theta',N)\,,
$
as we can check by an easy calculation. Its unitarity can be checked either by
a direct computation or by remarking that $U(A)$ is unitary on
${\Cal H}_z = N^2\cdot\int_0^1\int_0^1 d^2\theta\, \hzt$
and hence is also unitary on $\hzt$.
If $\theta$ has the property that $\beta(\theta)=\theta$, we can again define
the quantum propagator
  (up to a normalization factor) $V(A): \hzt\to \hzt$ by the formula
  $
  V(A,\theta, N)=D_\hbar P_{z,{{}\,{}^T\!\!A^{-1}}z}(\theta, N)\circ U(A),
 $
  which is the restriction of $V(A)$ in (2.8) to $\hzt$. This yields
exactly the same propagators as in \cite{DE}. A particularly simple
expression for $V(A,\theta,N)$ is obtained when $A$ is of the special form
considered in the following Corollary (see also \cite{HB}).

\proclaim{Corollary 3.1}
If $z=(1,0)$, $\theta=(0,0)$ and
$A=\left(\matrix 2g & 1\\ 4g^2-1 & 2g \endmatrix\right)\in SL(2,\Z)$,($g>1$,
$g\in\Z$) then:
$$
V(A,\theta, N)_{\ell,k}=\frac{1}{\sqrt{N}}
\exp{[\frac{2\pi i}{N}\,(g\ell^2-\ell k+gk^2)\,]}\,.
$$
\endproclaim

\demo{Proof of Proposition $3.2$}

1)  Let $v\in\Z^2$ with $\omega(w,v)=1$ and set $m=-Jw$, $\mt=Jv\,.$
We define, for $\teti\in\lbrack 0,1\lbrack\,\times\,\lbrack 0,1\lbrack$ ,
$$
S(\teti)=\sum_{\alpha,\beta\in\Z} (-1)^{N\alpha\beta}\, \exp{\lbrack\,-2\pi
i(\alpha\teti_1+\beta\teti_2)\,\rbrack}\, U(\alpha m+\beta\mt)\,.
$$
It is then easy to see that $S(\teti)$ is a continuous operator from ${\Cal
S}(\R^2)$
to $\Di$ which extends uniquely to a map from $\Di$ to $\Di$. Moreover, a
calculation
shows
$$
\align S(\teti)&\,=\,
\Bigl(\sum_{\alpha\in\Z} \exp[\,-2\pi i\teti_1\alpha\,]\,U(\alpha m)\Bigr)\,
\Bigl(\sum_{\beta\in\Z} \exp[\,-2\pi i\teti_2\beta\,]\,U(\beta\mt)\Bigr)\\
&{}\quad\quad\,=\,
\Bigl(\sum_{\beta\in\Z} \exp[\,-2\pi i\teti_2\beta\,]\,U(\beta\mt)\Bigr)\,
\Bigl(\sum_{\alpha\in\Z} \exp[\,-2\pi i\teti_1\alpha\,]\,U(\alpha m)\Bigr)
\endalign
$$
and
$$
U(\alpha' m+\beta '\mt)\,S(\teti)\,=\,(-1)^{N\alpha'\beta'}\,
\exp[\,2\pi i\,(\teti_1\alpha'+\teti_2\beta')\,]\,S(\teti)\,.
$$
Since
$$
\binom{\alpha'}{\beta '}=\left (\matrix v_1 & v_2\\ w_1 & w_2\endmatrix\right)
\binom{n_1}{n_2}
$$
for $\alpha' m+\beta '\mt=n$, we have
$$ u_1\,S(\teti)\,=\,(-1)^{Nv_1w_1}\,
\exp[\,2\pi i\,(v_1\teti_1+w_1\teti_2)\,]\,S(\teti)\,,
\tag{3.22}
$$
$$ u_2\,S(\teti)\,=\,(-1)^{Nv_2w_2}\,
\exp[\,2\pi i\,(v_2\teti_1+w_2\teti_2)\,]\,S(\teti)\,.
$$
As a result $S(\teti)\,\Di\subset \Df$, with
$$
\theta_1=(N/2) v_1w_1+\lbrack v_1\teti_1+w_1\teti_2\rbrack
\quad\quad\quad
\theta_2=(N/2) v_2w_2+\lbrack v_2\teti_1+w_2\teti_2\rbrack
\tag{3.23}
$$
For $\psi \in {\Cal D}_w$ as in (2.4), a simple computation using the
Poisson formula yields
$$
\lbrack \,S(\teti)\psi\,\rbrack(x)=\exp[\,-i\pi N\, h_w(x)h_v(x)\,]
\,\sum_{k\in\Z} d_k(\theta)\,
\delta[\,h_w(x)-(k+\teti_1)/N\,]
\tag{3.24}
$$
 where
$$  d_k(\theta)=\frac{1}{N}\,\sum_{\beta\in\Z} \exp[\,-2\pi
i\,\beta\teti_2\,]\,
f((k+\teti_1)/N-\beta)\,.\tag{3.25a}
$$
Note that
$$  d_{k+N}(\theta)=\exp[\,-2\pi i\,\teti_2\,]\, d_k\,.\tag{3.25b}$$
Using (3.23), one establishes
$$
(k +\teti_1)/N=\qkw-(1/2)\,w_1w_2\,[\,v_2 - v_1-1\,],\tag{3.26}
$$
with $\qkw$ as in (3.4), and
$$
\teti_2=-\alpha_{\theta}(N,w)-(N/2)\, v_1v_2\,[\,w_1 - w_2-1\,],
\tag{3.27}
$$
with $\alpha_{\theta}(N,w)$ defined in (3.6). Note that the relation
$w_1v_2-w_2v_1=1$ implies that the last term in (3.26) and in (3.27) is an
integer. Hence (3.24) becomes
$$
\lbrack \,S(\teti)\psi\,\rbrack(x)=\exp[\,-i\pi N\, h_w(x)h_v(x)\,]\,
\sum_{k\in\Z} c_k(\theta)\,\delta[h_w(x)-\qkw]
\tag{3.28}
$$
with $ c_k(\theta)=d_{k+\frac{N}{2}w_1w_2[v_2-v_1-1]}(\theta)$ satisfying
(3.5), thanks to (3.25) and (3.27). Hence (3.28) is written in the form (3.3)
which shows
$ S(\teti)\psi\in\dwt\,.
$
Recall now from (3.16) and (3.18) that $\hwt\cong \C^N$. As a result
$$
\align
N^2\,\int_0^1\int_0^1d^2\theta\,\hwt&\cong  N^2\,\int_0^1\int_0^1
\C^N\,d^2\theta\\
&\cong \,L^2(\,[0,1[\,\times\,
[0,1[\,;\,\C^N,\,N^2 d^2\theta).
\endalign
$$
On the other hand, if $f\in {\Cal S}(\R)\subset L^2(\R,dy)$, then, using
(3.25), and
performing a change of variables in the integral, using (3.23), yields
$$ \align
N^2\,\int_0^1 d\theta_1\int_0^1 d\theta_2&\,\sum_{r=0}^{N-1}
\vert c_r(\theta)\vert^2=\\
 &N^2\,\int_0^1 d\teti_1\int_0^1 d\teti_2\,\sum_{r=0}^{N-1}
\vert c_r(\theta)\vert^2=\int_{\R} \vert f(y)\vert^2\, dy.
\endalign
$$
Hence the map
$$
f\mapsto (c_0(\theta),\cdots,c_{N-1}(\theta))
\in L^2(\,[0,1[\,\times \,[0,1[\,; \C^N,\,
N^2 d^2\theta)
$$
extends to a natural isometry on all of $L^2(\R,dy)$. It is easily seen to be
onto
and hence unitary.
Since $L^2(\R,dy)\cong\hw$, this proves (1).
\newline 2) $\forall \psi_1\in\hw$, $\psi_2\in\hz$, we have
$$
\align \langle\psi_2, P_{zw}U(a,b)\psi_1\rangle_z
& =\int\frac{dq\,dp}{2\pi\hbar}\,\,
\overline{\psi_2}\,U(a,b)\psi_1 \\
& = \int\frac{dq\,dp}{2\pi\hbar}\,\,
\overline{U(a,b)^*\psi_2}\,\psi_1 =\,\langle U(a,b)^*\psi_2, P_{zw}\psi_1
\rangle_z.
\endalign
$$
3) We know from the remark after
Proposition 2.2 that $P_{zw}$ extends from $\hw$ to a map from $\dw$ to $\dz$.
  Moreover, in view of (2), $P_{zw} \dwt\subset\dzt$. Hence, defining
  ${\tilde P}_{zw}(\theta, N)$ to be the restriction of $P_{zw}$ to $\dwt$, it
follows that
  ${\tilde P}_{zw}(\theta, N):\hwt\to\hzt$ and consequently that
$P_{zw}= N^2\,\int d^2\theta\,
  {\tilde P}_{zw}(\theta, N)$.  By computing the
   explicit formula for ${\tilde P}_{zw}(\theta,N)$, we will show
$N\,{\tilde P}_{zw}(\theta,N)= P_{zw}(\theta, N)$, establishing the
proposition.
  Taking $\psi\in\dwt$ in the form (3.3), and using
  the definition of $P_{zw}$ in Proposition 2.2, we have
$$
\align  [\,&{\tilde P}_{zw}(\theta,N)\,\psi\,](x_2)=\\
&(1/2\pi\hbar\Delta)\,\exp[\,-i\pi N\,h_z(x_2)h_u(x_2)\,]\,
\sum_{k\in\Z}c_k \,\exp[\,-2\pi iN\,S_{zw}(q_k(w,\theta),h_z(x_2))\,]\,,
\tag{3.29}
\endalign
$$
where $S_{zw}$ is given in (2.7).
Letting $k=\ell\Delta N+r$, $\ell\in\Z$ and  $r\in\{0,\cdots,\Delta N-1\}\,,$
the {\it r.h.s.} of (3.29) reads
$$
\align
\exp[\,-i\pi N\,&h_z(x_2)h_u(x_2)\,]\,\cdot\\
 \frac{N}{\Delta}\,&\sum_{\ell\in\Z}\sum_{r=0}^{N\Delta-1} c_r\,
\exp[\,2\pi i\,\ell\Delta\alpha_{\theta}(N,w)\,]\,\exp[\,-2\pi i
N\,S_{zw}(q_r+\ell\Delta,h_z(x_2))\,]\,.
\endalign
$$
Since $(1/2)\,\omega(v,z)\,\ell^2\Delta=(1/2)\,\omega(v,z)\,\ell\Delta\,\,
\mod 1\,,$  we have
$$
S_{zw}(q_r+\ell\Delta,h_z(x_2))
=S_{zw}(q_r,h_z(x_2))+\ell\,[q_r\omega(v,z)+h_z(x_2)]+(1/2)\omega(v,z)
\ell^2\Delta\,,
$$
so that
$$
\align
\exp[\,&-2\pi iN\,S_{zw}(q_r+\ell\Delta,h_z(x_2))\,]=\\
{}&\exp[\,-2\pi i N\,S_{zw}(q_r,h_z(x_2))\,]\,
\exp[\,-2\pi i N\,(q_r\omega(v,z)+h_z(x_2)+\omega(v,z)\Delta/2)\ell\,]\,.
\endalign
$$
This yields:
$$
\align
[\,&{\tilde P}_{zw}(\theta,N)\psi\,](x_2)=
\exp[\,-i\pi N\,h_z(x_2)h_u(x_2)\,]\,\cdot\\
{} & \Bigl(\frac{N}{\Delta}\,\sum_{r=0}^{N\Delta-1}c_r\,\exp[\,-2\pi i
N\,S_{zw}(q_r,h_z(x_2))\,]\Bigr)\,\Bigl(\sum_{\ell\in\Z}
\exp[\,-2\pi iN\,(h_z(x_2)-A)\ell\,]\Bigr)\,,
\endalign
$$
where
$$  A=-q_r(w,\theta)\,\omega(v,z)-(1/2)\,\omega(v,z)\Delta+(\Delta/N)\,
\alpha_{\theta}(N,w),
$$  and where $q_r(w,\theta)$ and $\alpha_{\theta}(N,w)$ are given by (3.4)
and (3.6) respectively. Note that we can replace $A$ by anything else mod
$1$, a freedom we will use in order to get a convenient form for
${\tilde P}_{zw}(\theta,N)$. In particular one has
$$
\align
A&=-(r/N)\,\omega(v,z)+(1/2)\,z_1z_2-(1/N)\,\omega(\theta,z)\quad\mod 1\\
&=q_{[-r\omega(v,z)]}(z,\theta)\quad\mod 1\,,
\endalign
$$
so that
$$
\align  [&{\tilde P}_{zw}(\theta,N)\,\psi](x_2)=
\exp[\,-i\pi N\,h_z(x_2)h_u(x_2)\,]\cdot\\
&{}\quad\quad\Bigl(\frac{N}{\Delta}\,\sum_{r=0}^{N\Delta-1}c_r\,
\exp[\,-2\pi iN\,S_{zw}(q_r,h_z(x_2))\,]\Bigr)\,
\Bigl((1/N)\,\sum_{k\in\Z}\delta
[y_2-q_{k-r\omega(v,z)}]\Bigr)\\
&=\exp[\,-i\pi N\,h_z(x_2)h_u(x_2)\,]\\
&{}\frac{1}{\Delta}\quad\sum_{r=0}^{N\Delta-1}c_r\,\sum_{k\in\Z}
\exp[\,-2\pi iN\,S_{zw}(q_r(w,\theta),
q_{k+r\omega(z,v)}(z,\theta))\,]
\,\delta[\,y_2-q_{k+r\omega(z,v)}(z,\theta)\,]\\
&=\exp[\,-i\pi N\,h_z(x_2)h_u(x_2)\,]\\
&{}\frac{1}{\Delta}\quad\sum_{r=0}^{N\Delta-1}c_r\,\sum_{\ell\in\Z}
\exp[\,-2\pi iN\,S_{zw}(q_r(w,\theta),
q_{\ell}(z,\theta))\,]
\,\delta[\,h_z(x_2)-q_\ell(z,\theta)\,]\,.
\endalign
$$
Using (3.5) and comparing to (3.20), one sees that
$\tilde P_{zw}(\theta, N)= (1/N)\,
P_{zw}(\theta, N)$. Hence $P_{zw}= N\int d^2\theta\, P_{zw}(\theta, N)$.\qed
\enddemo
To summarize, by applying the ideas of geometric quantization in their simplest
form, one can easily quantize linear transformations on $\R^2$ as well as on
$\T$. We stress again that the construction is simple and calculationally
very convenient. Indeed, although the proofs of Propositions 3.1 and 3.2 are
somewhat involved in the general case, they reduce to trivialities when
$\Delta =1$, as in Corollary 3.1 and in the following sections.  In that case
(3.20)
does not involve a sum and the unitarity of $P_{zw}$ is then immediate.
We shall now show that the reformulation of geometric quantization we have just
presented allows for an immediate generalization to a class of piecewise linear
or
affine linear transformations of the torus.

\bigskip
\bigskip

\subheading{4. Quantization of piecewise linear and affine transformations}\par
\bigskip
\noindent {\sl $($A$)$ Translations and skew translations.}
\medskip
The simplest transformations on the torus are undoubtedly the translations
$x=(q, p)\mapsto(q+a, p+b)\,\mod 1\,$. If $a={r_1}/{s_1}$ and $b={r_2}/{s_2}$
(with $g.c.d.$ $(r_i, s_i)=1$, $i=1,2\,$), then we can write
$(a,b)=(r/s)\,(-w_2,w_1)$ for integer $r,s$ with $g.c.d.$ $(r, s)=1$,
$w\in\Z^2$, and  $g.c.d.$ $(w_1, w_2)=1$. Here $s$ is the least common
multiple of $s_1$ and $s_2$,
which is also the common period of all orbits under this translation.

Taking $k\in\N^*$, $N=sk$, we saw in Section 3 (see (3.17)) how to quantize
this  translation. The expression of the
quantum translation $U(a,b)$ (i.e. (3.17) with $\ell=rk$) shows that its
eigenfunctions
are concentrated on the circles
$$\omega(x,(a,b))=(r/s)\,q_i\qquad i=0,\dots, ks-1$$ and that they are $k$-fold
degenerate.   The quantum propagator is easily seen to have the same
period as the classical translation since
$$U^s(a,b) = \exp [\,2\pi i\,(-\frac{w_1w_2}{2}\,\ell s
+ r\omega(w,\theta))\,]\,\, id_{\hwt}.$$
It follows that, as in the multidimensional harmonic oscillator with
commensurate frequencies \cite{DBIH}, these degeneracies can be used to
construct eigenfunctions of $U(a,b)$ that, in the
classical limit ($k\to\infty$), concentrate on any given classical orbit.

The approach of Section 3 does not a priori permit the quantization of
translations of the form $(a,b)=\alpha\,({r_1}/{s_1},{r_2}/{s_2}),\,\,
\alpha\notin\Q$, much less of ergodic translations, for
which $a/b\notin\Q$.  The reason is that the corresponding prequantized
translations
do not preserve the spaces
$\hwt$.

Since the ergodic translations are undoubtedly the simplest ergodic
dynamical systems, it would be interesting to circumvent this difficulty and
to nevertheless construct a quantum analog for them.  We will see that
this can be done very naturally within the framework of  Section 3.
The situation is actually very similar to the one encountered when  quantizing
linear flows. Indeed, there we saw that
$U(A)\hwt={\Cal H}_{{}\,^T\!\!A^{-1}w}(\theta, N)$
for a suitable choice of $\theta$ and then we used the natural pairing between
${\Cal H}_{{}\,^T\!\!A^{-1}w}(\theta, N)$ and $\hwt$ to construct $V(A)$. Here
we will see that $U(a,b)\hwt=\hwtp$ with $\theta'$ given in Lemma 4.1 below.
Although
in this case
we can never choose $\theta$ so that $\theta'=\theta$, we will construct an
 identification
$D_{\hbar} P_{vw}(\theta,\theta')$ between $\hwtp$ and $\hvt$ in analogy with
(3.19).
Since there is also a natural identification
$D_{\hbar}P_{wv} (\theta)$ between $\hvt$ and $\hwt$ (Proposition 3.1), we
define
the unitary quantum translation $Q_w(a,b)$ by $$
Q_w(a,b)=D_{\hbar}^2 \,P_{wv} (\theta,\theta)\circ P_{vw}(\theta,\theta')\circ
U(a,b)\,:\,\hwt\to\hwt\,.\tag{4.1}
$$
Note that this reduces to (3.17) when the translation has the required form,
and that
the $Q_w(a,b)$ depend continuously on $(a,b)$.  On the other hand, the
construction is
$w$-dependent and it is clear that the $Q_w(a,b)$ can not generate a unitary
representation of the full Weyl-Heisenberg group.

\proclaim{Lemma 4.1}
\roster
\item $U(a,b)\,\Df=\Dfp\,$, with
$
(\theta_1',\theta_2')=(\theta_1 -Nb,\theta_2 +Na)\quad \mod 1\,.
$
\item $U(a,b)\nabla_w \psi=\nabla_w U(a,b)\psi\,$
for any $w\in\R^2$, $(a,b)\in\R^2$, $\psi\in\Di\,$.
\item $U(a,b):\hwt\to\hwtp$ is unitary.
\endroster
\endproclaim
\demo{Proof}
Both (1) and (2) follow from a simple computation. That $U(a,b)$  maps
$\hwt$ isomorphically onto $\hwtp$ is an immediate consequence of (1) and (2).
To check the unitarity, let $\psi\in\hwt$ with
$$
\psi(q,p)=\sum_{k\in\Z} c_k\,\exp[\,{-i\pi N\, h_w(x) h_v(x)}\,]\,
\delta\bigl(h_w(x)-q_k(\theta,w)\bigr)\,.
\tag{4.2}
$$
For convenience, we write $\tau=(\tau_1,\tau_2)=(a,b)$. Now we introduce
$\ell=(\ell_1,\ell_2)\in\Z^2\,$, $I_{1/N}=\,]-{1\over N},{1\over N}\,[\,$
and $\beta=(\beta_1,\beta_2)\in\,I_{1/N}^2\,$, uniquely determined by
$$
\align
\tau_i&={\ell_i}/{N}+\beta_i\\
\theta'_i&=\theta_i+(-)^i\,N\,\beta_{3-i}\,\in\,[\,0,1\,[\,,
\endalign
$$
with $i=1,2$. A direct calculation shows that
$$
[\,U(a,b)\psi\,](q,p)=\sum_{k\in\Z} d_k\,\exp[\,{-i\pi N\,h_w(x) h_v(x)}\,]
\,\delta\bigl(h_w(x)-q_k(\theta',w)\bigr)\,,
$$
where
$$
d_k=c_{k-h_w(\ell)}\,
\exp[\,{i\pi N\,\bigl(2 q_k(\theta',w)h_v(\tau)-h_w(\tau)h_v(\tau)\bigr)}\,]\,
\tag{4.3}
$$
Recalling the identification $\psi\cong (c_0,\cdots,c_{N-1})$ and
$[U(a,b)\psi]\cong (d_0,\cdots,d_{N-1})$, the unitarity of $U(a,b)$ is
now immediate from (4.2).\qed
\enddemo

Given now $U(a,b)\psi\cong (d_0,\cdots,d_{N-1})\in {\Cal H}_w(\theta ',N)$, we
can
 proceed in the spirit of Proposition 3.1 to define $P_{vw}(\theta,\theta'):
{\Cal H}_w(\theta', N)\to\hvt$ as follows:
$$\langle\psi_2, P_{vw}(\theta, \theta')\psi_1\rangle_{\hvt}=
\int_{[0,1)\times [0,1)}\,{\overline\psi_2}\psi_1\,
{dq\,dp\over2\pi\hbar}.\tag{4.4}$$
A simple calculation then yields
$$[P_{vw}(\theta,\theta')\psi](q,p)=N\,\sum_{\ell}[P_{vw}(\theta,\theta')\psi]_{\ell}\,
\exp{[i\pi Nh_w(x)h_v(x)]}\,\delta(h_v(x)-q_{\ell}(v,\theta))
$$
where
$$
\align
[P_{vw}(\theta,\theta')\psi]_{\ell}&=N\, \sum_{k=0}^{N-1} d_k\,\exp{[-2i\pi N
S_{vw}
(q_k(w,\theta'),q_{\ell}(v,\theta))]}\\
&=N\, \sum_{k=0}^{N-1}d_k\,\exp{[-2i\pi N q_k(w,\theta')q_{\ell}(v,\theta)]}
\endalign
$$
It is easy to see that $\parallel D_{\hbar}P_{vw}(\theta,\theta')
U(a,b)\psi\parallel_{{\Cal H}_v (\theta,N)}^2=\parallel \psi\parallel_{{\Cal
H}_w(\theta
',N)}^2$, where $\vert D_{\hbar}\vert =N^{-3/2}$.

When $(a,b)$ is ergodic, the eigenfunctions of the $Q_w(a,b)$ can on general
grounds be expected to be equidistributed on the torus in the classical limit,
in sharp contrast to what happens in the periodic case.

Note that it is now easy to quantize skew translations of the form
$(q,p)\mapsto(q+a, p+kq)$ which are
ergodic if $a$ is irrational and $k$ a non-zero integer \cite{CFS}. They are
just the
composition of a linear transformation and a translation.
\bigskip
\noindent {\sl $($B$)$ Piecewise affine transformations.}
\medskip
A first class of piecewise affine maps studied in \cite{Ch} is the following.
Take $A=\left(\matrix a & b\\ c & d \endmatrix\right)\in SL(2,\Z)$, apply it to
$[ 0,1)\times [ 0,1)$, then cut the resulting parallelogram into
strips along the direction $(a,c)$ and shift the strips around with
translations
parallel to $(a,c)$. Combining Section 3 and Section 4A, one can easily obtain
a
quantization for this class of transformations.

\medskip
Let us now turn to another class of discontinuous maps described in
\cite{Ch,LW,V}.
Consider the map $A_1=\left(\matrix 1 & b\\ 0 & 1\endmatrix\right)$, $b\in\R$
restricted to the strip $0\leq p\leq 1$ and taken modulo $1$ in $q$. This
defines a map $A_1$
on the torus, discontinuous on the circle $\{p\in\Z\}$ if $b\notin\Z$.
Similarly, construct a map $A_2$ on the torus by restricting $A_2=\left(\matrix
1 & 0\\ b &
1\endmatrix\right)$, $b\in\R$ to the strip $0\leq q\leq 1$ and taking $p$
modulo $1$.
This map will be discontinuous on the circle $\{q\in\Z\}$ if $b\notin\Z$.
The map $A=A_2A_1$, which is a discontinuous
hyperbolic area preserving map on the torus, is ergodic and exponentially
mixing, \cite{Ch,Li,LW,V}.

We now propose a quantization of $A_i$, $i=1,2$ in the spirit of Section 3.
Calling $V_i$ the quantization of $A_i$, we will define the quantum propagator
$V$ of
$A$ by $
V=V_2V_1$.
We saw in Section 2 that $U(A_1)\dw={\Cal D}_{(A_1^{-1})^T w}$. If, however,
$a\notin\Z$
then $U(A_1)\Df\not\subset\Dfp$ for any choice of $\theta$ and $\theta '$.
This situation is similar to, but slightly more complicated than, the one of
the
previous paragraph, where $U(a,b)\dw=\dw$, but $U(a,b)\Df=\Dfp$.
So there is again no geometrically natural definition of the quantum propagator
associated to $A_1$.
This reflects the fact that $A_1$ is not a  continuous automorphism of the
torus.
The approach of Section 3 nevertheless suggests an obvious way to quantize
$A_1$.
For that purpose, note that the image of $\lbrack 0,1)\times\lbrack 0,1)$ under
$A_1$ is
$$
F_1=\{ (q,p)\in\R^2\,\vert\,0\leq p<1,\,\,bp\leq q<bp+1\},
$$
which is again a fundamental domain of the torus.
Let $w=(1,0)$, $v=(0,1)$. Then, if $\psi\in\hwt$ and $\varphi\in\hvt$,
it is immediately
clear, because of the transversality of the lines $p=p_k$, $q+bp=q_{\ell}$,
that
$\overline{\varphi} U(A_1)\psi$ still defines a distribution on the plane.

As a result, there exists a unique map $PU(A_1):\hwt\to\hvt$ defined by
$$
\langle\varphi,PU(A_1)\psi\rangle_{\hvt}=\int_{F_1}
\overline{\varphi}U(A_1)\psi\,\frac{dq\,dp}{2\pi\hbar}.
\tag{4.5}
$$
Here the right-hand side of ($4.5$) is to understood as the value of the
distribution
$\overline{\varphi} U(A_1)\psi$ on a smooth characteristic function of $F_1$.
Explicitly, a simple calculation shows that, for any $k,\ell=0,\cdots, N-1$
$$
\lbrack PU(A_1)\rbrack_{k\ell}=N\exp{[-i\pi N b p_k^2]}\,\exp{[-2i\pi N
q_{\ell}p_k]}
$$
where $q_{\ell}=\ell/N +\theta_2/N$, $p_k=k/N-\theta_1/N$.

The resulting quantum propagator on $\hwt$ is then, using the natural
identification
between $\hvt$ and $\hwt$:
$$
V_1=N^{-3/2}{\Cal F}_N^{-1}\circ PU(A_1).
$$
Note that $N^{-3/2} PU(A_1)$ itself is the product of the finite Fourier
transform
with the diagonal matrix $D_1$ with entries $\exp{[-i\pi N b p_k^2]}$.
So
$$
V_1={\Cal F}_N^{-1}\circ D_1\circ {\Cal F}_N.\tag{4.6}
$$
Remark that for $b\in\Z$ and for the appropriate $\theta$ this reduces to the
result
obtained in Section 3, as is easily checked.
Note furthermore that the map $A_1$ behaves as a completely integrable
transformation with
invariant circles $p=const.$ This is perfectly reflected in the structure of
$V_1$.
{}From equation ($4.6$) one sees that its eigenfunctions are indeed
concentrated on the
invariant circles.

Finally, the construction of $V_2$ is completely analogous, with the roles of
$w$
and $v$ interchanged.
The resulting quantum propagator $V=V_2V_1$ on $\hwt$ is readily seen to be
$$
V=D_2\circ{\Cal F}_N^{-1}\circ D_1\circ {\Cal F}_N.\tag{4.7}
$$
Here $D_2$ is the diagonal matrix with entries $\exp{[i\pi N b q_{\ell}^2]}$.
The non trivial structure of $V$ comes from the fact that it is the product
of two non commuting matrices $V_1, V_2$.
\bigskip
\noindent {\sl $($C$)$ The Baker transformation.}
\medskip
Given the matrix $A=\left(\matrix 2 & 0\\ 0 & 1/2\endmatrix \right)$, we
consider the
piecewise affine map $B$ defined on the unit square ($0\leq q < 1, 0\leq p\leq
1)$ by
$$
B(q,p)=\cases
Ax\,,\phantom{T(-1,1/2)\circ A}\quad 0\leq q< 1/2\,,\\
T(-1,1/2)\circ A\,,\phantom{Ax}\quad 1/2\leq q<1\,, \endcases
$$
where $T(a,b)x=(q+a,p+b)$.
This map is called {\it  the Baker transformation},
and its dynamical properties have been studied in detail (see \cite{AA,LW}).
Note that it has the same structure as the piecewise affine maps described
above. First one applies a linear map, then one slices the resulting rectangle
and shift the parts around. There is one major difference, however, leading to
some
additional complications for the quantization. The linear part of the Baker
transformation
does not send $\lbrack 0,1)\times\lbrack 0,1)$ into another fundamental domain
of $\T$.

Even though the Baker transformation is not continuous on the torus, the tools
we developed
in the previous section
 can again be used to associate a corresponding quantum
operator to this map, as we now show.  In particular, as in \cite{BV, Sa},
we take the point of view that the correct quantum Hilbert spaces for this
problem are
still the ones constructed in Section 3 (see below).
It then suffices to mimic the approach of the previous section, with some
minor changes to account for the discontinuities of the map.
The resulting quantum operator is identical to the one obtained in \cite{BV,Sa}
by
completely different arguments.

We shall first define a prequantized version ${\QB}$ of $B$ on distributions on
$\R^2$
with support in the left or right half of the unit square.
Suppose $\psi$ is a distribution supported on $0\leq q<{1\over 2}\,$,
${}\,\,0\leq p\leq 1\,$. Then we define
$$
(\QB\psi)(q,p)=U(A)\,\psi(q,p)\,.
$$
Note that the support of $\QB\psi$ is contained in $0\leq q<1\,$,
${}\,\,0\leq p\leq {1\over 2}\,$. If, on the other hand, $\psi$ is supported in
${1\over 2}\leq q<1\,$, ${}\,0\leq p\leq 1$, then
$$
(\QB\psi)(q,p)=[\,U(-1,1/2)\circ U(A)\,]\psi(q,p)
$$
and its support is now contained in $0\leq q<1\,$, ${}\,\,{1\over 2}\leq p\leq
1$.

Given $N\in \N$, and $w=(1,0)$, recall that $\dwt$ is the space of
distributions $\psi$
of the form:
$$
\psi(q,p)=\sum_{k\in\Z}c_k\,\exp[\,{-i\pi  N\,pq}\,]\,\delta(q-q_k)\,,
$$
where $q_k=k/N +\theta_2/N $ and, in addition,
$c_{k+N}=e^{-2\pi i\,\theta_1}\,c_k$ for any $k\in\Z\,.$ Therefore, because of
the latter relations, no information is lost if we restrict $\psi$ to the
unit square, namely
$$
\psi(q,p)=\sum_{k=0}^{N-1} c_k\, \exp[\,{-i\pi N\,pq}\,]\,
\delta(q-k/N-\theta_2/N) \,\,\chi_{[0,1]}(p)\,,\tag{4.8}
$$
where $\chi_{[0,1]}$ is the characteristic function of the unit interval.  We
shall
write $H_1(\theta)$ for the space of distributions of the form (4.8), equipped
with the
inner product (3.18).  This is the quantum Hilbert space for the Baker map in
the {\sl
position representation}, which is realized as a space of distributions on the
phase
space of the problem.  Similarly, we introduce  $H_2(\theta)$, which is the
space of
distributions ${\Cal D}_v(\theta, N)$ with $v=(0,1)$, restricted to the unit
square, i.e.,
$\psi\in
H_2(\theta)$ iff
$$\psi = \sum_{\ell=0}^{N-1} d_l\exp{[i\pi
Npq]}\,\delta(p-p_l)\,\chi_{[0,1]}(q),$$
where $p_\ell =\ell/N + \theta_1/N$.  $H_2(\theta)$ is the quantum
Hilbert space of the Baker transformation in the {\sl momentum} representation.
 We have
a natural identification between $H_1(\theta)$ and $H_2(\theta)$, given by the
pairing of section 3, which in this case is just the finite Fourier transform
(see the remark after Proposition 3.1).

We now observe that we have a natural decomposition $H_1(\theta)=H_L(\theta)
\bigoplus H_R(\theta)$.
Indeed, each $\psi\in H_1(\theta)$ can
be uniquely written as $\psi=\psi_L + \psi_R$, where
$$
\align
\psi_L&=\sum_{0\leq q_k < {1\over 2}} c_k\,\exp{[\,-i\pi
N\,pq\,]}\,\delta(q-q_k) \,\,
\chi_{[0,1]}(p)\,,\\
\psi_R&=\sum_{{1\over 2}\leq q_k<1} c_k\,\exp{[\,-i\pi N\,pq\,]}\,\delta(q-q_k)
\,\,
\chi_{[0,1]}(p)\,,\\
\endalign
$$
have their respective supports in $0\leq q<{1\over 2}$, and
${1\over 2}\leq q<1$. We can now compute
$$
(\QB\psi)(q,p)=(\QB\psi_L)(q,p)+(\QB\psi_R)(q,p)\,.
$$
This gives
$$
\align
(\QB\psi_L)(q,p)&=2\sum_{0\leq q_k <{1\over 2}} c_k\,\exp[\,{-2\pi
iN\,q_kp}\,]\,
\delta(q - 2q_k)\,\chi_{[0,1]}(2p)\,,\\
(\QB\psi_R)(q,p)&=2\, \exp[\,{2\pi i\, (\theta_2-N/4)\,]}\,
\sum_{{1\over 2}\leq q_k < 1}\,c_k\,\exp[\,{-2\pi iN\,(q_k-1/2)\,p}\,]\,\cdot\\
{}&\phantom{2\, \exp[\,{2\pi i\, (\theta_2-\frac{N}{4})\,]}\,
\sum_{k=N/2}^{N-1}\,c_k\,\quad\quad} \delta(q+1-2q_k)\,\,\chi_{[0,1]}(2p-1)\,.
\endalign
$$
Note that the support of $\QB\psi_L$ is contained in $0\leq q<1,\,
 0\leq p\leq {1 \over 2}$, whereas the support of $\QB\psi_R$ is contained in
$0\leq q<1,\,{1\over 2}\leq p\leq 1$. It is clear that $\QB\psi$ obtained in
this way is
 not an element of $H_1(\theta)$ (for
any $\theta$), nor of
any $\dzt$.  Hence, we have no hope of applying the general results on pairing
of
the previous section directly to define the quantum propagator.  It will
nevertheless
turn out that we can again define, in the spirit of (3.19), a natural
projection $P\hat
B\psi$ of the
distribution $\hat B\psi$ onto $H_2(\theta)$.
\proclaim{Proposition 4.1}  If $N$ is even and $0<\theta_1,\,\theta_2<1$, then
there exists a
unitary map $2^{-1/2}N^{-3/2}\,P\QB:H_1(\theta)
\to H_2(\theta)$, uniquely defined by:
$$\langle\psi_2,P\hat B\psi_1\rangle_{H_2(\theta)}=\int_{[0,1)\times [0,1)}
\overline\psi_2 \hat B\psi_1 \,
{dq\,dp\over 2\pi\hbar}.\tag{4.9}$$
Specifically,
$$
P(\QB\psi)(q,p)=\sum_{\ell=0}^{N-1} \bigl(P(\QB\psi)\bigr)_\ell\,
\exp[\,{i\pi N\, qp}\,]\,\delta(p-p_\ell)\,\chi_{[0,1]}(q),
$$
with, for $\ell<N/2$,
$$
\bigl(P(\QB\psi)\bigr)_\ell=2N\,\sum_{k=0}^{N/2-1}\,c_k\,
\exp[\,{-4\pi iN \,q_kp_\ell}\,]\,,
$$
and for $\ell\geq N/2$,
$$
\bigl(P(\QB\psi)\bigr)_\ell=2N\,\,\exp[\,{2\pi i\, (\theta_2-N/4)\,]}\,
\sum_{k=N/2}^{N-1} c_k\,\exp[\,{-4\pi iN\,((q_k-1/2)\,p_\ell)\,]}\,.
$$
\endproclaim
\remark{Remark} We omit the proof, obtained by a simple computation.  Let us
point
out that the conditions on $\theta$ assure that $\overline\psi_2 \hat B\psi_1$
is a distribution on the plane, with support in the unit square.  They
guarantee in
particular that
$\psi_2$ does not have support on the line $p=1/2$, which would lead to
technical problems involving the multiplication with $P\QB\,\psi_1$. The right
hand
side of (4.9) is to be understood as the value of this distribution on a smooth
characteristic function of the unit square, and is independent of its choice.
The
unitarity statement does not follow from the results of section 3, since
$P\QB\,\psi_1$
is not a polarized section, as pointed out above.  In this sense, the unitarity
of
the construction is somewhat surprising.  It breaks down when $N$ is odd,
although the
block diagonal structure of $P\QB$ would permit to restore it by hand.
\endremark

We now define the quantum Baker  transformation $V_B$ in the spirit of section
3
 as follows:
$$V_B= 2^{-1/2}N^{-3/2}\,{\Cal F}_N^{-1}\circ P\QB:H_1(\theta)\to
H_1(\theta),$$
where we used the natural pairing between $H_2(\theta)$ and $H_1(\theta)$
described
above. A simple calculation now shows that if $\theta=(0,0)$ $V_B$ is
exactly the operator
obtained in \cite{BV}. If $\theta=(1/2,1/2)$,
$V_B$ coincides with the quantum Baker map of \cite{Sa}. Although the value
$\theta=0$
is strictly speaking excluded by the Proposition, it can be obtained in the
limit.
We mention that this construction can be immediately extended
to a more general class of Baker like transformations \cite{BV}.

In conclusion, these examples show that the framework of section 3 permits the
treatment of situations that are not geometrically natural and would therefore
not
be tractable within the framework of geometric quantization as such.  Remark
for example
that, although the right hand side of equation (4.9) makes sense, it is not
geometrically
intrinsic, unlike the right hand side of (3.19).  Similarly, the identification
of the
 quantum  Hilbert spaces with $\C^N$ in section 3 was merely a calculational
devise, which
is again no longer the case here. Nevertheless, it is clear that the phase
space
formulation of quantum mechanics given by geometric quantization automatically
reproduces the clever intuitive arguments used to construct the quantized Baker
transformation in \cite{BV}. In particular, the prequantized map is very close
to
the classical map: this is clear from the general expression for
$\exp[\,-{i\over\hbar}\,]\hat f t$ in
Section 2.  As a result, it still has the "left to bottom", "right to top"
structure of the classical map. In \cite{BV} this feature was built into the
construction of the quantized Baker transformation by assumption .



\Refs
\refstyle{A}
\widestnumber\key{BNSGG}
\ref\key AA\by V.I. Arnold and A. Avez\book Ergodic Problems of Classical
Mechanics\publ Benjamin,New York\yr 1968\endref
\ref\key Be\by F.A. Berezin\paper General concept of quantization
\jour Commun. Math. Phys.\vol 40 \pages 153--174 \yr 1975\endref
\ref\key Bl\by R.J. Blattner\paper Quantization and Representation
Theory\pages 147-161 \jour Proc. Sympos. Pure Math.\vol 26\yr 1973\endref
\ref\key BV\by N.L. Balazs, A. Voros\paper The Quantized Baker's Transformation
\jour Annals of Physics\vol 190\pages 1-31\yr 1989\endref
\ref\key CFS\by I.P. Cornfeld, S.V. Fomin and Ya. G. Sinai \book Ergodic
theory\publ
Springer Verlag\publaddr Berlin\yr 1982\endref
\ref\key Ch \by N. Chernoff\paper Ergodic and statistical properties of
piecewise linear hyperbolic automorphisms of the two-torus
\pages 111--134 \jour J. Stat. Phys. \vol 69 \yr  1992 \endref
\ref\key DBHI \by S. De Bi\`{e}vre, J.C. Houard and M. Irac
\paper Wave packets localized on closed classical trajectories
\jour in Differential Equations with Applications to Mathematical Physics,
Eds. W.F. Ames, E.M. Harrell, J.V. Herod, Academic Press Inc. Boston
\yr 1993\endref
\ref\key DE\by M. Degli Esposti\paper Quantization of the orientation
preserving
automorphisms of the torus\jour Ann. Inst. H. Poincar\'e\vol 58\pages 323-341
\yr 1993\endref
\ref\key DGI\by M. Degli Esposti, S. Graffi and S. Isola\paper Classical limit
of
the quantized hyperbolic toral automorphisms
\jour to appear in Commun. Math. Phys.\yr 1994\endref
\ref\key Eck \by B. Eckhardt \paper Exact eigenfunctions  for a quantized map
\jour J. Phys. A\vol 19 \yr 1986 \pages 1823--1833\endref
\ref\key Fo \by G. Folland \book Harmonic Analysis in Phase Space
\publ Princeton University Press \publaddr Princeton \yr 1988\endref
\ref\key GuSt\by V. Guillemin and S. Sternberg\book Geometric Asymptotics
\publ Mathematical Surveys\vol 14\yr 1977\endref
\ref\key HB \by J.H. Hannay, M.V. Berry \paper Quantization of  linear maps on
a torus - Fresnel diffraction by a periodic grating
\pages 267--291 \jour Physica D\vol 1\yr 1980\endref
\ref\key Ke1 \by J. Keating \jour Ph.D. thesis University of Bristol\yr 1989
\endref
\ref\key Ke2 \by J. Keating \paper Asymptotic properties of the periodic orbits
of the cat maps \pages 277--307 \jour Nonlinearity
\vol 4 \yr  1991 \endref
\ref\key Ke3 \by J. Keating \paper  The cat maps: quantum mechanics and
classical motion \pages 309--341 \jour Nonlinearity \vol 4 \yr  1991 \endref
\ref\key Ko\by B. Kostant\paper Quantization and Unitary Representations
\pages 87-208 \jour Lecture Notes in Math.\vol 170\yr 1970\endref
\ref\key Li \by C. Liverani \paper  Decay of correlations
\jour To be published in Annals of Math. \yr  1994 \endref
\ref\key LW\by C. Liverani, M.P. Wojtkowski\paper Ergodicity in Hamiltonian
systems\jour to appear in Dynamics Reported\endref
\ref\key Sa\by M. Saraceno\paper Classical Structures in the Quantized Baker
Transformation\jour Annals of Physics\vol 199\pages 37-60\yr 1990\endref
\ref\key Sar\by P. Sarnak \paper Arithmetic Quantum Chaos \inbook Tel Aviv
Lectures 1993
\endref
\ref\key Sn\by J. Sniatycki\book Geometric Quantization and Quantum Mechanics
\publ Applied Mathematical Sciences\vol 30\yr 1980\endref
\ref\key V \by S. Vaienti \paper Ergodic properties of the discontinuous
sawtooth map \pages 251 \jour J. Stat. Phys. \vol 67 \yr  1992 \endref
\ref\key Wo\by N.M.J. Woodhouse\book Geometric Quantization\publaddr Clarendon
Press, Oxford \yr 1980\endref

\endRefs
\enddocument